\documentclass{aip-cp}

\usepackage[numbers]{natbib}
\usepackage{rotating}
\usepackage{graphicx}

\newcommand{\Erho}{ \left[755(2)(1)(^{20}_{02})-\frac{i}{2}\,129(3)(1)(^{7}_{1})\right]~{\rm MeV}}
\usepackage{color}

\usepackage{subfigure}
\usepackage[caption = false]{subfig}

\begin{document}

\title{From QCD to Physical Resonances}

\author[cu,bay]{Daniel R. Bolton\corref{cor1}}
\author[jlab,odu]{Ra\'ul A. Brice\~no}
\author[camb]{David J. Wilson}

\affil[cu]{Department of Physics, University of Colorado, Boulder, CO 80309, USA}
\affil[bay]{Department of Physics, Baylor University, Waco, TX 76798, USA}
\affil[jlab]{Thomas Jefferson National Accelerator Facility, 12000 Jefferson Avenue, Newport News, VA 23606, USA}
\affil[odu]{Department of Physics, Old Dominion University, Norfolk, VA 23529, USA}
\affil[camb]{Department of Applied Mathematics and Theoretical Physics, Centre for Mathematical Sciences, University of Cambridge, Wilberforce Road, Cambridge, CB3 0WA, UK}

\corresp[cor1]{Corresponding author: daniel.bolton@colorado.edu}

\maketitle

\begin{abstract}
In this talk, we present the first chiral extrapolation of a resonant scattering amplitude obtained from lattice QCD. Finite-volume spectra, determined by the \emph{Hadron Spectrum Collaboration} at $m_\pi=236$~MeV~\cite{Wilson:2015dqa}, for the isotriplet $\pi\pi$ channel are analyzed using the L\"uscher method to determine the infinite-volume scattering amplitude. Unitarized Chiral Perturbation Theory is then used to extrapolate the scattering amplitude to the physical light quark masses. The viability of this procedure is demonstrated by its agreement with the experimentally determined scattering phase shift up to center-of-mass energies of 1.2~GeV. Finally, we analytically continue the amplitude to the complex plane to obtain the $\rho$-pole at $\Erho$.
\end{abstract}

\section{INTRODUCTION}

A major goal of nuclear physics is a deep understanding of the non-perturbative nature of Quantum Chromodynamics (QCD). One manifestation of this non-perturbative nature is the spectrum of hadronic resonances. Perhaps the most well known and understood hadronic resonance is the $\rho$ meson. Because it is a resonance, all properties of the $\rho$ must be deduced from the behavior of the hadronic system that produces the $\rho$, namely the $(I,J)=(1,1)$ $\pi\pi$ system. In this talk we review a recent determination of the $\pi\pi$ scattering amplitude presented in Ref.~\cite{Bolton:2015psa}. This determination was accomplished by using effective field theory (EFT) to extrapolate the lattice QCD results presented in Ref.~\cite{Wilson:2015dqa} by the \emph{Hadron Spectrum Collaboration} using quark masses that correspond to $m_\pi=236$~MeV.
 
Lattice QCD, with which one can compute correlation functions directly from QCD in a discrete and finite spacetime, has proven to be a great tool for understanding many aspects of QCD. Yet few-body systems continue to present great difficulties in large part due to the finite-volume nature of the calculation. Neither resonances nor asymptotic states exist in a finite-volume. Nevertheless, one can access information regarding asymptotic states from finite-volume effects (we point the reader to Ref.~\cite{Briceno:2014tqa} for a recent review). The non-perturbative mapping between finite-volume energy levels and infinite-volume scattering observables was first derived in Refs.~\cite{Luscher:1986pf, Luscher:1990ux} by Martin L\"uscher. Since then, many generalizations have extended the applicability of this ``L\"uscher method" for all two-body systems~\cite{Rummukainen:1995vs, Kim:2005gf, Christ:2005gi, He:2005ey, Briceno:2012yi, Hansen:2012tf,  Briceno:2014oea}.

Generalizing the methodology presented by L\"uscher for systems where three particles or more can simultaneously go on-shell has proven to be challenging.
\footnote{In spite of this there has been a great deal of progress in the literature~\cite{Hansen:2015zta, Hansen:2014eka, Hansen:2015zga, Briceno:2012rv, Polejaeva:2012ut}, which gives hope that the quantization condition for generic systems composed of at most three-particles will be derived in the upcoming years.}
Even if a quantization condition describing few-body systems in a finite volume were known, its application would require a formidable effort. To understand this, it helps to consider the aforementioned example of the $\rho$-resonance. In nature, this is a state that lies above the $\pi\pi$ and $4\pi$ thresholds and these two channels can couple. In a finite volume, this would lead to a mixing between these channels, and as a result, there would no longer be a one-to-one mapping between the spectrum and $\pi\pi$ scattering.  For two-body systems, this mixing between different scattering channels is a well-understood phenomenon~\cite{He:2005ey, Briceno:2012yi, Hansen:2012tf, Briceno:2014oea}, and it is a challenge that, at least in practice, has been surmounted by the \emph{Hadron Spectrum Collaboration} \cite{Dudek:2014qha, Wilson:2014cna, Wilson:2015dqa}. 

For the time-being, it is advantageous to perform lattice QCD calculations of resonances using unphysically heavy quark-masses, moving the few-body thresholds to higher energies. This, in part, led the \emph{Hadron Spectrum Collaboration} to calculate the $(I,J)=(1,1)$ $\pi\pi$ spectrum using at $m_\pi=236$~MeV. This assures that the $4\pi$, $6\pi$, and $K\overline{K}$ thresholds lie well above the $\rho$-resonance, allowing one to use the L\"usher method to extract the $\pi\pi$ scattering amplitude and subsequently the $\rho$ pole. Performing lattice QCD calculations at unphysically heavy quark masses introduces the need for an extrapolation towards the physical point before a comparison with experiment can be made. In this work, we use Unitarized Chiral Perturbation Theory (U$\chi$PT)~\cite{Oller:1997ng, Dobado:1996ps, Oller:1998hw, GomezNicola:2001as, Pelaez:2006nj}.

 The $\rho$-resonance has been extensively studied using $N_f=2+1$~\cite{Wilson:2015dqa, Metivet:2014bga, Dudek:2012xn, Bulava:2015qjz, Aoki:2011yj} and $N_f=2$~\cite{Bali:2015gji, Guo:2015dde, Pelissier:2012pi, Lang:2011mn, Feng:2010es, Aoki:2007rd} dynamical quarks,
\footnote{For a more complete review of spectroscopy efforts from lattice QCD, we point the reader to Lang's review talk presented in this same conference~\cite{Lang:2015ljt}.}
 and U$\chi$PT has been be previously advocated as a tool to extrapolate LQCD results~\cite{Chen:2012rp, Doring:2012eu, Doring:2011nd, Doring:2011vk, Bernard:2010fp, Nebreda:2011di, Rios:2008zr, Hanhart:2008mx, Pelaez:2010fj}. Our work represents the first extrapolation of a resonant scattering amplitude. We demonstrate that the use of these methods in this channel are sufficient for an accurate post-diction of experimental $\pi\pi$ elastic scattering up to center-of-mass energies of 1.2 GeV.

\begin{figure}[t]
\centering
{\includegraphics[scale=0.35]{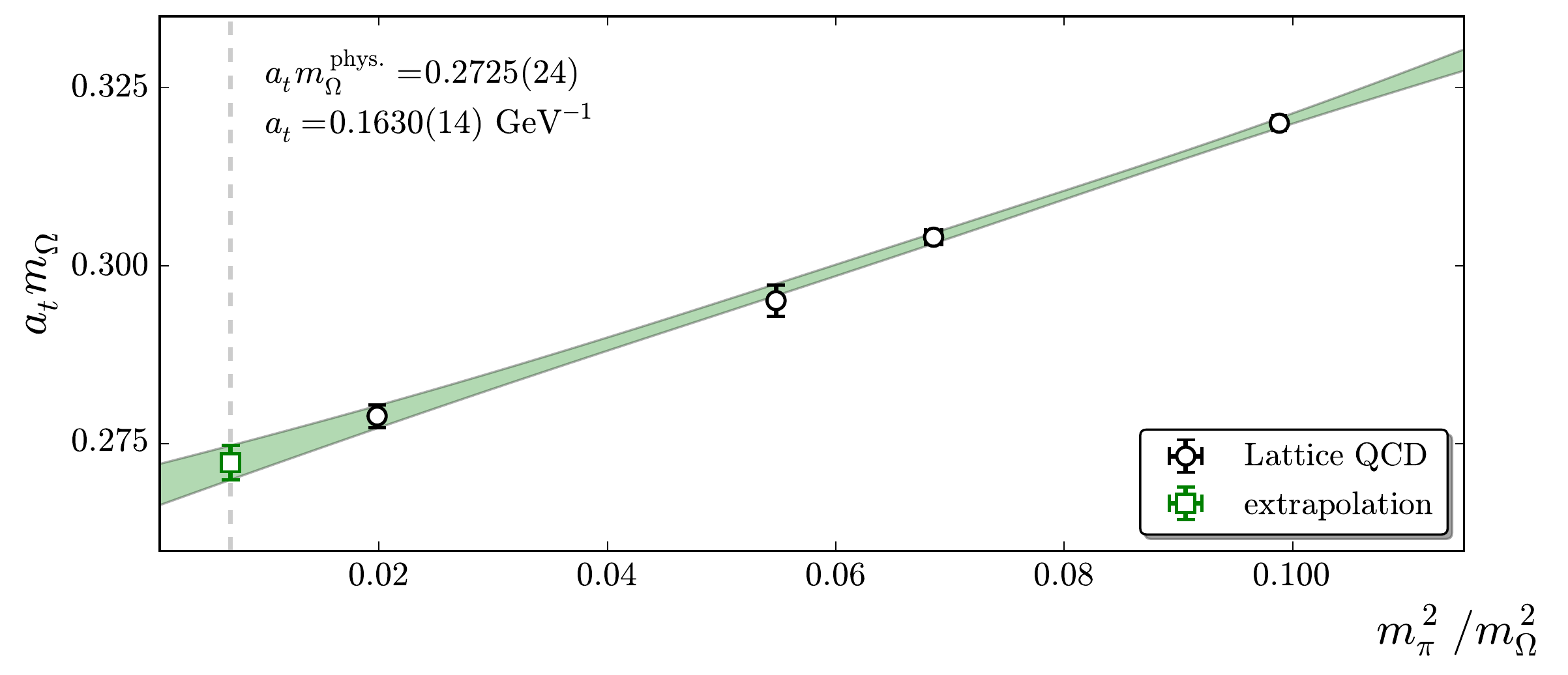}}
\hspace{.5cm}
{\includegraphics[scale=0.35]{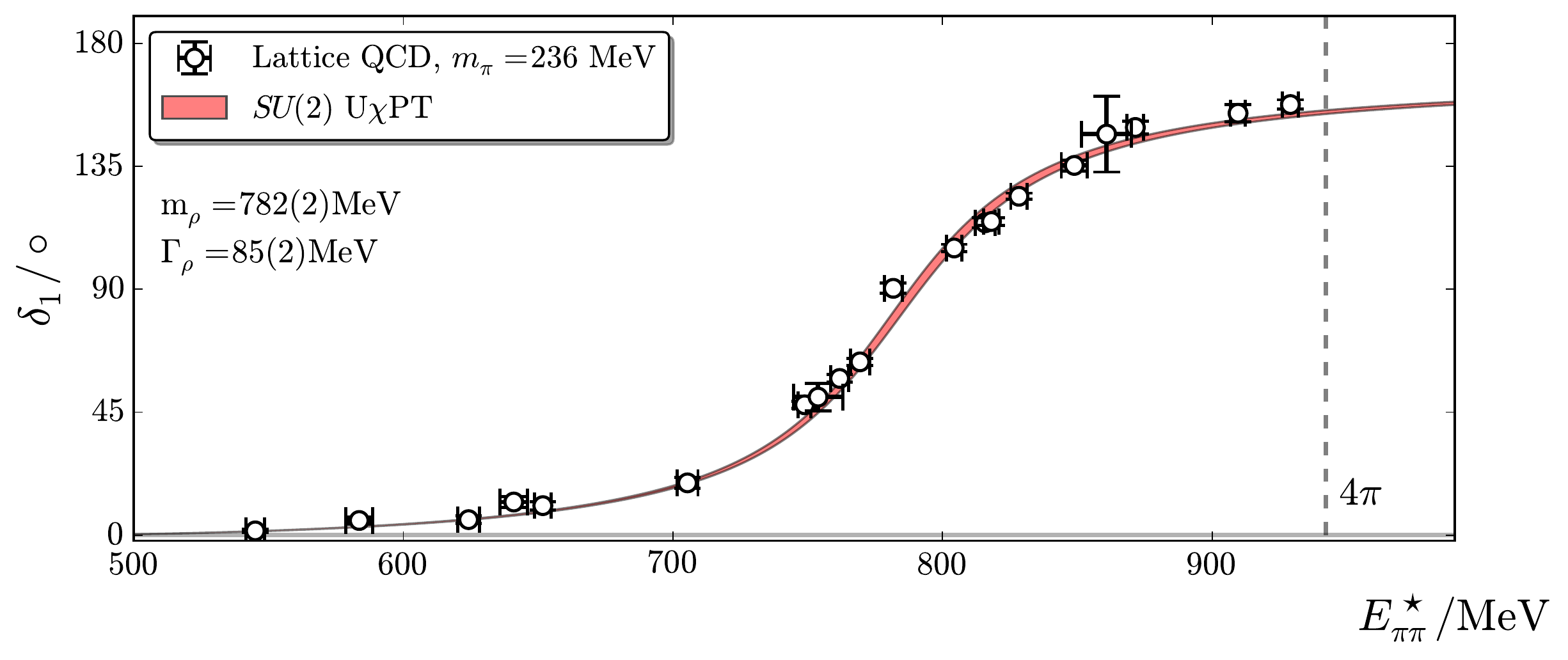}}
\caption{On the left: Shown is the chiral extrapolation of $a_t \Omega$ (with points from Refs.~\cite{Lin:2008pr, Wilson:2015dqa}). On the right: The $(I,J)=(1,1)$ $\pi\pi$ scattering phase shift obtained from a fit using the SU(2) U$\chi$PT scattering amplitude (red curve) to the discrete points obtained in Ref.~\cite{Wilson:2015dqa}.}
\label{fig:m_omega_lattice_fits}
\end{figure}

\section{METHOD AND RESULTS}
As outlined in the introduction, we need only use a single-channel version of the L\"usher method to relate the infinite-volume scattering amplitude $\mathcal M(P)$ to this particular finite-volume spectrum. The quantization condition for two-particles with arbitrary total momenta in a cubic volume can be written as~\cite{Rummukainen:1995vs, Kim:2005gf, Christ:2005gi},
\begin{equation}
\label{eq:QC}
\det[F^{-1}(P,L) + \mathcal M(P)] = 0\,,
\end{equation}
where $F(P,L)$ is a known function that depends on the total momentum ${\bf P}$ and the box size $L$, $\mathcal M$ is the infinite volume scattering amplitude, and the determinant acts in the space of partial waves. Corrections to this relation are exponentially suppressed by the relatively large value of $m_\pi L\approx4.4$~\cite{Wilson:2015dqa}.

To parametrize the scattering amplitude we use \emph{unitarized chiral perturbation theory} (U$\chi$PT)~\cite{Oller:1997ng, Oller:1998hw, GomezNicola:2001as}, which is derived from standard $\chi$PT~\cite{Weinberg:1966kf, Colangelo:2001df,Ecker:1988te,Gasser:1984gg,Gasser:1983yg,Gasser:1983kx} by applying the Inverse Amplitude Method. There are several reasons why one might use U$\chi$PT instead of standard $\chi$PT. U$\chi$PT satisfies unitarity order-by-order, and as a result can be used to study bound states and resonances.
\footnote{In this sense, the amplitudes obtained via U$\chi$PT closely resemble the resulting amplitude obtained in nucleon-nucleon systems using the KSW expansion~\cite{Kaplan:1998we, Kaplan:1998tg}.}
This procedure has been shown to accurately describe low-lying resonances in multiple channels with a relatively small set of parameters (see, for example, Refs.~\cite{Oller:1997ng, Oller:1998hw, GomezNicola:2001as}). Furthermore, it extends the kinematic range of applicability of standard $\chi$PT. 

In the present work, we focus our attention on SU(2) $\chi$PT/U$\chi$PT, which uses the nearly exact isospin symmetry in nature (an exact symmetry in most lattice calculations). The three flavor SU(3) theory suffers from relatively large systematic errors. At each order in the chiral expansion, $\chi$PT/U$\chi$PT require a finite number of parameters that need to be fixed. At leading order (LO) in the chiral expansion, there are two ``low energy coefficients" (LECs)  $m_0, f_0$. These can be fixed by the pion mass and decay constant respectively. At next-to-leading order (NLO) in the chiral expansion, four additional LECs appear in the Lagrangian ($\ell^r_1,\ell^r_2,\ell^r_3,\ell^r_4$)~\cite{Gasser:1983kx}. The superscript $r$ indicates the renormalized value of a LEC. In this work, we use $\mu=770$~MeV as the renormalization scale. These LECs can be fixed using the scattering amplitude. In the $(I,J)=(1,1)$ scattering amplitude, only two linear combinations appear,
\begin{equation}
\alpha_1 \equiv -2\ell^r_1+\ell^r_2, \hspace{2cm}
\alpha_2 \equiv \ell^r_4.
\end{equation}
Although the next-to-next-to-leading order scattering amplitudes are known~\cite{Bijnens:1995yn}, we only use the expression up to NLO. The U$\chi$PT scattering amplitude, $\mathcal{M}_{\rm U\chi PT}$, can be written in the terms of the standard LO and NLO scattering amplitudes, $\mathcal{M}_{\rm LO}$ and $\mathcal{M}_{\rm NLO}$,~\cite{Oller:1997ng, Oller:1998hw, GomezNicola:2001as},
\begin{equation}
\mathcal{M}_{\rm U\chi PT}
=\mathcal{M}_{\rm LO}\frac{1}{\mathcal{M}_{\rm LO}-\mathcal{M}_{\rm NLO}}
\mathcal{M}_{\rm LO}.
\label{eq:amp}
\end{equation}

\begin{figure}[t]
{\includegraphics[scale=0.35]{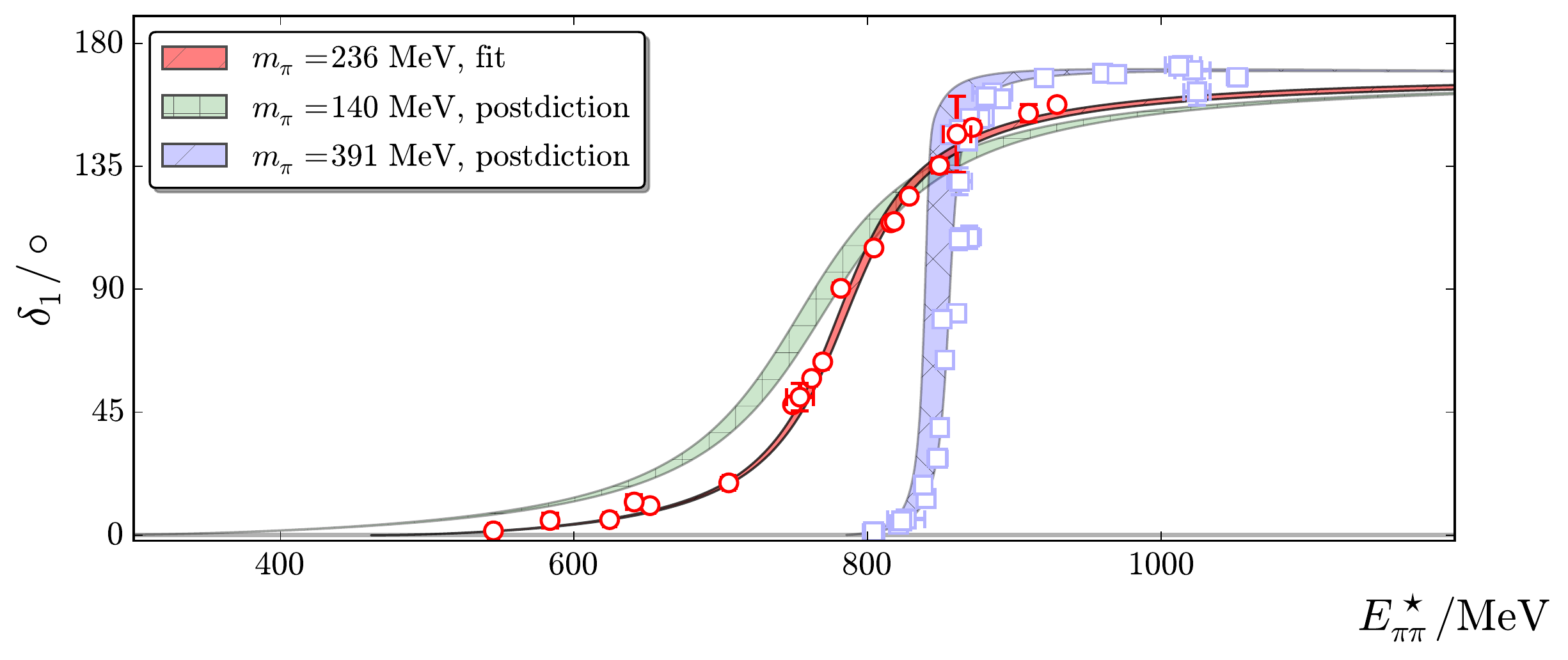}}
\hspace{.5cm}
{\includegraphics[scale=0.35]{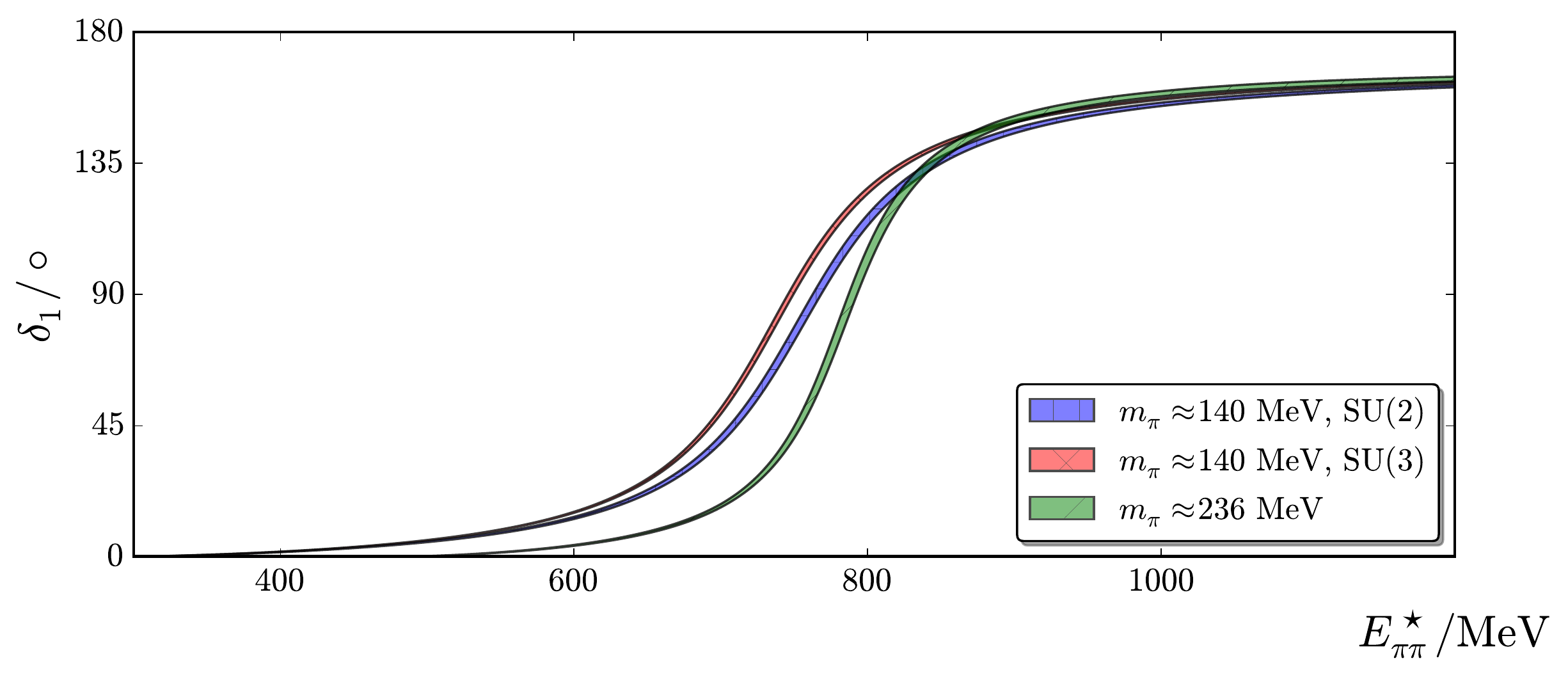}}

\caption{On the left: The elastic $\pi\pi$ scattering phase shift shown in Fig.~\ref{fig:m_omega_lattice_fits} (red), extrapolated to higher (blue) and lower (green) values of $m_\pi$ using SU(2) U$\chi$PT. The discrete blue points are from Ref.~\cite{Dudek:2012xn} while the red points are repeated from Fig.~\ref{fig:m_omega_lattice_fits}. On the right: Extrapolations from $m_\pi=236$ MeV to $m_\pi=140$ MeV using the SU(2) (blue) and SU(3) (red) versions of U$\chi$PT.}
\label{fig:extrapolations}
\end{figure}

Having the unitarized amplitude at our disposal, we may proceed to fit the $\pi\pi$ spectrum to determine $\alpha_{1,2}$. Reference~\cite{Wilson:2015dqa} provided 22 energy levels (in a single volume) with which to perform the fit. As a measure of the goodness of fit, we defined 
\begin{equation}
\chi^2(\{\alpha_i\})=\sum_{j,k}
\delta{E}_{j}(\{\alpha_i\})\,
{\mathbb C}^{-1}_{j,k}\,
\delta{E}_{k}(\{\alpha_i\})
\end{equation}
where $\delta E_j(\{\alpha_i\})=\left[E_j^{\rm lat}-E_j^{{\rm U}\chi{\rm PT}}({\alpha_i})\right]$, the indicies $\{j,k\}$ run over all 22 energy levels, and ${\mathbb C}$ is the covariance matrix.

Once the LECs are determined, one has a scattering amplitude that can be applied at any value of $m_\pi$ (subject to convergence of U$\chi$PT). Before doing this, we must determine the lattice spacing. Reference~\cite{Wilson:2015dqa} chose to define the scale using the omega baryon, which has a mild dependence on $m_\pi$. More explicitly, since $a_tm_\Omega$ was determined in Ref.~\cite{Wilson:2015dqa} one can set $m_\Omega$ equal to its physical value to obtain $a_t$. Alternatively, one can use a simple $m_\pi$-dependence,
\begin{equation}
\label{eq:m_omega}
m_\Omega(m_\pi)=m_{\Omega,0}+\alpha \frac{m_\pi^2}{m_\Omega^2}+\beta \frac{m_\pi^4}{m_\Omega^4}
\end{equation}
and fit the previously determined $a_tm_\Omega$ for a range of values of the quark masses~\cite{Lin:2008pr, Wilson:2015dqa}. This requires assuming that the lattice spacing is the same for all these ensembles. One can extrapolate to obtain $a_t$ as shown in the left panel of Fig.~\ref{fig:m_omega_lattice_fits}. We use the result of this procedure to estimate the systematic uncertainty in our results due to the determination of the lattice spacing.

The fit resulted in the following LECs with a $\chi^2$ per degree of freedom of 1.26,
\begin{equation}
\begin{array}{l}
\alpha_1(770\ {\rm MeV})=14.7(4)(2)(1)\times10^{-3}\\
\alpha_2(770\ {\rm MeV})=-28(6)(3)\left(^{01}_{11}\right)\times10^{-3}
\end{array}
\quad
\left[
\begin{array}{lr}
1 & -0.98\\
 & 1
\end{array}
\right]
\label{eq:lecs}
\end{equation}
where the first uncertainty is statistical, the second is the systematic due to the determinations of $m_\pi$ and $\xi$ (the anisotropy of the lattice), and the third is the systematic due to the determination of the lattice spacing. The symmetric matrix on the right of the coefficients denotes the statistical correlation between the two. See the right panel of Fig.~\ref{fig:m_omega_lattice_fits} for a plot of the scattering phase shift (at $m_\pi=236$~MeV) using these extracted LECs. We note that when this amplitude is analytically continued to the complex plane to extract the pole of the resonance we find $E_\rho=782(2)-\frac{i}{2}85(2)$~MeV. This is in good agreement with the poles obtained by the \emph{Hadron Spectrum Collaboration}, which used a different set of parametrizations for the scattering amplitude.

Figure~\ref{fig:extrapolations} shows the results of using these extracted LECs to compute the scattering phase shift at other values of $m_\pi$. The left panel shows the theory at $m_\pi=140, 236,$ and $391$~MeV. The $m_\pi=391$~MeV theory is compared with a prior calculation by the \emph{Hadron Spectrum Collaboration}~\cite{Dudek:2012xn}. It is worth nothing that although U$\chi$PT should not be trusted for such a heavy value of the pion mass, one finds qualitatively good agreement with the lattice QCD calculation. The right panel compares the results of extrapolating to the physical point with SU(2) vs. SU(3) U$\chi$PT. As previously mentioned, SU(3) $\chi$PT suffers from relatively large systematic errors that we are not able to properly estimate. 
 
\begin{figure}[t]
\centerline{\includegraphics[scale=0.53]{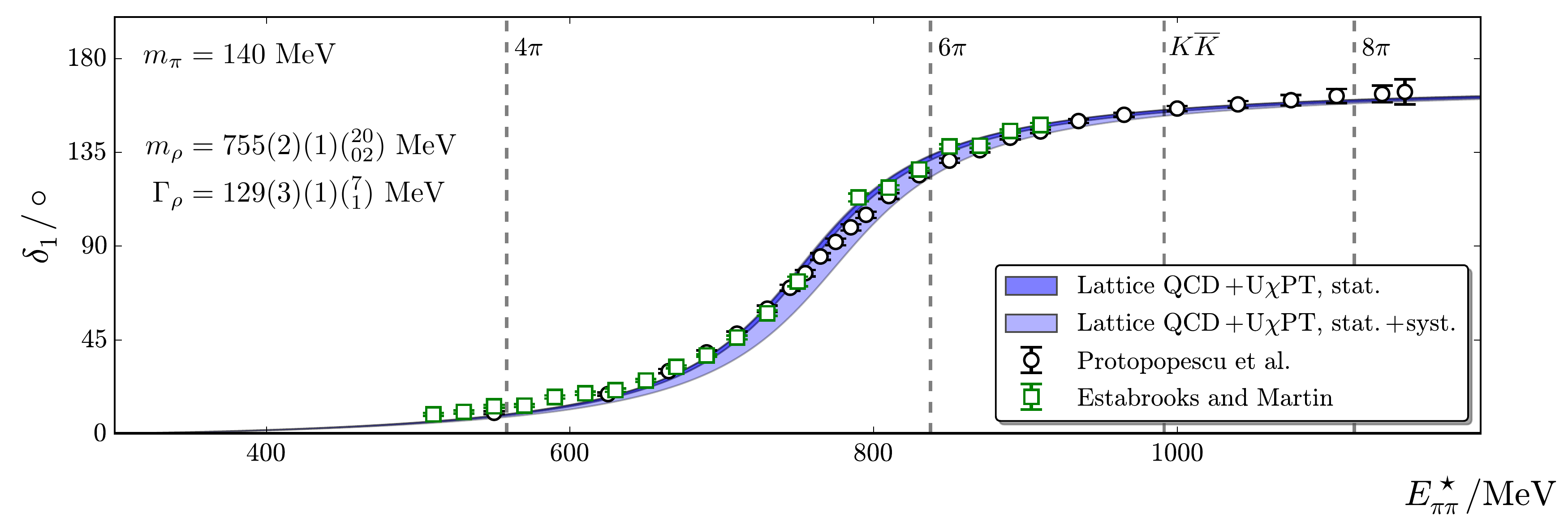}}
\caption{The extrapolated resonant $(I,J)=(1,1)$ $\pi\pi$ scattering phase shift from lattice QCD and U$\chi$PT shown alongside experimental data from Refs.~\cite{Protopopescu:1973sh,Estabrooks:1974vu}. the dark blue band encompasses the statistical error only, while the light blue band includes systematic errors discussed in the text. The dashed lines denote the $4\pi, 6\pi,  K\overline{K},$ and $8\pi$ thresholds. }
\label{fig:global_comparison}
\end{figure}

Finally, in Fig.~\ref{fig:global_comparison} we present our extrapolated phase shift in comparison with experimental data~\cite{Protopopescu:1973sh,Estabrooks:1974vu}. We observe good agreement for points up to center of mass energy ($E^\star_{\pi\pi}$) of 1.2 GeV, well beyond the point at which standard $\chi$PT would fail. Furthermore, it is important to note that our result reinforces the fact that the few-body thresholds, namely those for the $4\pi,\, 6\pi,\,  K\overline{K},$ and $8\pi$ states, seem to give negligible contributions. This is emphasized by the dashed lines in the plot denoting the thresholds. By analytically continuing the amplitude to the complex plane, we find the pole at $E^\star_{\pi\pi}=\Erho$.
 
\section{CONCLUSION AND OUTLOOK}

The study of few-body systems via lattice QCD continues to pose a challenge. This has restricted studies of resonant amplitudes to unphysically heavy quark masses, where few-body thresholds can be safely neglected. To connect lattice QCD determinations of the resonant isotriplet $\pi\pi$ scattering amplitude, we present a chiral extrapolation of the results by the \emph{Hadron Spectrum Collaboration}~\cite{Wilson:2015dqa}. This demonstrates that, in practice, U$\chi$PT can be used for extrapolation of resonant amplitudes determined from lattice QCD using moderately light quark masses.

There are two obviously desirable classes of systems to consider in the future. First are the more phenomenologically interesting isoscalar channels, where lattice QCD calculations are still at their early stages~\cite{Dudek:2013yja, Dudek:2011tt}. The second are the study of electromagnetic processes involving resonant states, where there has been exciting progress~\cite{Briceno:2015dca, Shultz:2015pfa, Feng:2014gba, Bulava:2015qjz}.

\section{ACKNOWLEDGMENTS}
We thank our colleagues in the Hadron Spectrum Collaboration, in particular J. J. Dudek,  R. G. Edwards and C.~E.~Thomas, for providing access to the lattice QCD spectrum and for useful discussions and feedback on the manuscript. D.R.B. would like to thank J. Emerick, C. Madrid, and K. Robertson for their help with this project. 
\nocite{*}
\bibliographystyle{aipnum-cp}%
\bibliography{bibi}%

\end{document}